\documentclass[reqno]{article}
\usepackage{alltt}
\usepackage{graphicx}
\title{The role of the $CP(N-1)$ geometry in the intrinsic unification of the general
relativity and QFT}
\author{Peter Leifer}
\date{Crimea State Medical University, Simferopol, Ukraine}
\begin{document}
\maketitle

\begin{abstract}
Einstein's program of the unified field theory transformed nowadays to the TOE requiring new primordial elements and relations between them. Definitely, they must be elements of the quantum nature. One of most fundamental quantum elements are pure quantum states. Their basic relations are defined by the geometry of the complex projective Hilbert space. In the framework of such geometry all physical concepts should be formulated and derived in the natural way.  Analysis following this logic shows that inertia and inertial forces are originated not in space-time but it the space of quantum states since they are generated by the deformation of quantum states as a reaction on an external interaction
or self-interaction. In particular, inertia law generalized by Einstein during development of general relativity (GR) will be expressed in intrinsic quantum terms. It is assumed that quantum formulation of the inertia law should clarify the old problem of inertial mass (dynamical mass generation). The conservation of energy-momentum following form this
law has been applied to self-interacting quantum Dirac's electron.
\end{abstract}
\vskip 0.1cm
\section{Introduction}
The second quantization method is the basis of the QFT \cite{Dirac_QFT}; it presents the top of the linear approach to the essentially non-linear problem of quantum interaction and self-interaction. This formal apparatus realizes the physical concept of the corpuscles-wave duality. It is well known, however, that this universality should be broken for interacting quantum fields \cite{Bl1}. Divergences and necessity of the renormalization procedure during solution of the typical problems of QED is the most acute consequence of this method \cite{Dirac_QFT}. These problems stem partly from the formal, ``stiff" method of quantization being applied to the dynamical variables and to the Fourier components of wave function. However, ideally, the quantization might be realized by the soliton-like lumps currying discrete portions of the mass, charge, spin, etc.
Such a way requires new non-linear wave equations derived from unknown ``first principles" connected with unified theory of quantum interactions. Modern attempts to unify initially electromagnetic, weak and strong interactions are based upon obvious observation that gravity is much weaker than other fundamental forces. But such approach is similar to splashing a water together with a kid since these attempts close the way to understand the source of inertia and gravity - a mass.

I would like discuss here intrinsic unification of general relativity
and non-linear quantum dynamics based on the eigen-dynamics of pure quantum state in
$CP(N-1)$ leading to the geometric formulation of non-Abelian gauge fields $\vec{P}=P^i(x,\pi)\frac{\partial}{\partial \pi^i} + c.c. = P^{\mu}(x)\Phi^i_{\mu}(\pi)\frac{\partial}{\partial \pi^i} + c.c.$
carrying the self-interaction \cite{Le1,Le2,LeMa1}. These gauge fields are considerably
differ from the Yang-Mills fields $\hat{A}_{\mu}(x)=A_{\mu}^{\alpha}(x)\hat{\lambda}_{\alpha}$
since from the technical point of view the fixed matrices $\hat{\lambda}_{\alpha} \in AlgSU(N)$ have been replaced by the smooth state-dependent vector fields on $CP(N-1)$.
Principally, in the base of such approach lies idea of the priority of internal $SU(N)$ symmetry and its breakdown with subsequent classification of quantum motions in $CP(N-1)$
over space-time symmetries \cite{Le3,Le4,Le5}. This idea leads to the
the quantum formulation of the inertia principle in the space state of the internal degrees of freedom, not in space-time \cite{Le1}. Namely, such ``matter field" like electron is represented by a dynamical process of the geodesic motion in $CP(N-1)$ and the geodesic variations shall be related to the gauge fields. Then the first order quasi-linear PDE's field equations for electron itself taking the place of the boundary conditions for the variation problem and Jacobi fields play the role of the generalized affine non-Abelian state-dependent gauge field associated with the gauge group $H=U(1) \times U(N)$. How one may identify this field with electroweak field will be discussed elsewhere. I will be concentrated here on the problem of the rest field mass of self-interacting electron and the origin of the inertial forces. In order to do this I will discuss initially motivation leading to new formulation of the inertia principle.

\textbf{Clarification }

0. The Einstein's convention of the summation in identical co- and contra-variant indexes
has been used.

1. Flexible quantum setup (FQS) is anholonomic reference frame $A^{\mu}(x)\Phi^i_{\mu}(\pi) \frac{\partial }{\partial \pi^i}$ in $CP(N-1)$ whose space-time coefficient functions $A^{\mu}(x)$ realizing a quantum setup ``tuning"  by variation of these components.

2. Local dynamical variables (LDV's) are vector fields
$\Phi^i_{\mu}(\pi) \frac{\partial}{\partial \pi^i} + c.c.$ on $CP(N-1)$ corresponding to the $SU(N)$ generators \cite{Le4}.

3. Superrelativity means physical equivalence of any conceivable quantum setup, i.e.
the quantum numbers of ``elementary" particles like mass, charge, spin, etc.,
are the same. This invariance grantees the self-identity
of quantum particles in any ambient.

4. Self-identity means the conservation of the fundamental LDV's corresponding mentioned quantum numbers.

5. The conservation of LDV's may be expressed by the affine parallel transport of LDV's in $CP(N-1)$.

6. Coefficient functions $\Phi^i_{\mu}(\pi)$ of the $SU(N)$ generators acting on quantum states in $CP(N-1)$ replace classical force vector fields acting on material point with charge, mass, etc. Being multiplied and contracted with potentials $P^{\mu}(x)$ they comprise vector field of proper energy-momentum giving the rate of the quantum state variation $\frac{d \pi^i}{d \tau}=\frac{c}{\hbar}P^{\mu}\Phi^i_{\mu}(\pi)$.
Their divergency $L_{\lambda}= \frac{\partial
\Phi_{\lambda}^n (\pi)}{\partial \pi^n} + \Gamma^m_{mn} \Phi_{\lambda}^n(\pi)$ may be treated as non-Abelian charges. From the formal point of view it comes out that the projective Hilbert space $CP(N-1)$ serves as base of the principle fiber bundle with $SU(N)$ structure group instead of the space-time with Poincare group and its representations as in the traditional QFT.

7. No connections of this theory with the commonly used sigma $CP(N-1)$ models in
low-dimensional space-time. $CP(N-1)$ compact manifold serves as base manifold;
4D space-time arises in the frame fibre bundle.

\section{Intrinsic unification of relativity and quantum principles}
The localization problem in space-time mentioned above and deep difficulties of divergences in quantum field theory (QFT) insist to find a new primordial quantum element instead of the classical material point and its probabilistic quantum counterpart. I will use unlocated quantum state of a system - a specific quantum motion \cite{Dirac}
as such primordial element. Quantum states of single quantum particles may be represented by vectors $|\Psi>, |\Phi>,...$ of linear functional Hilbert space $\mathcal{H}$ with countable or even finite dimensions since these states
related to the internal degrees of freedom like spin,charge, etc. It is important
to note that the correspondence between quantum state and its vectors representation in $\mathcal{H}$ is not isomorphic. It is rather homomorphic, when a full equivalence class of proportional vectors, so-called rays $\{\Psi\}= z |\Psi>$, where $z \in \mathcal{C} \setminus \{0\}$ corresponds to the one quantum state $|\Psi>$. The rays of quantum states may be represented by points of complex projective Hilbert space $CP^{\infty}$ or its finite dimension subspace $CP(N-1)$. Points of $CP(N-1)$ represent generalized coherent states (GCS) that will be used thereafter as fundamental physical concept instead of material point. This space will be treated as the \emph{space of ``unlocated quantum states"} as the analog of the \emph{``space of unlocated shapes"} \cite{SW}.
We will dealing with the lift of the quantum dynamics from $CP(N-1)$ into the \emph{space of located quantum states}. That is, the variance between Shapere \& Wilczek construction and our scheme is that the dynamics of unlocated quantum states should be represented by the motions of the localizable 4D ``field-shell" in dynamical space-time.

Two simple observations serve as the basis of the intrinsic unification of relativity and quantum principles. The first observation concerns interference of quantum states in a fixed quantum setup.

A. The linear interference of pure quantum states (amplitudes) shows the symmetries
relative space-time transformations of whole setup. This interference has
been studied in ``standard" quantum theory. Such symmetries reflects, say,
the \emph{first order of relativity}: the physics is same if any
\emph{complete setup} subject (kinematical,
not dynamical!) shifts, rotations, boosts as whole in a single Minkowski
space-time. According to our notes given some times ago \cite{Le1,Le6} one should add to this list a freely falling quantum setup (super-relativity).

The second observation concerns a dynamical ``deformation" of some quantum setup.

B. If one dynamically changes the setup configuration or its ``environment", then the pure state (amplitude of an event) will be generally changed. Nevertheless there is a different type of tacitly assumed deep symmetry that may be formulated on the intuitive
level as the invariance of physical properties of ``quantum particles",
i.e. the invariance of their quantum numbers like mass, spin, charge,
etc., relative variation of the quantum state due to the ambient variance. This means that the physical properties expressed by intrinsic dynamical variables of, say, electrons in two different setups $S_1$ and $S_2$ are the same. It is close to the nice Fock's idea
of ``relativity to observation devices" but it will be realized in infinitesimal form as following:

\emph{One postulates that the invariant content of this physical properties may be kept if one makes the infinitesimal variation of some
``flexible quantum setup"  reached by a small variation of some
fields by adjustment of tuning devices.}

 A new concept of local dynamical variable (LDV) \cite{Le5} should be introduced for the realization of the ``flexible quantum setup". This construction is naturally connected with methods developed in studying geometric phase \cite{Berry,Littlejohn}. I seek, however, conservation laws for LDV's in the quantum state space following from the new
formulation of the inertia principle.

\section{Quantum formulation of the inertia principle}
Success of Newton's conception of physical force
influencing on a separated body may be explained by the fact that the \emph{geometric counterpart to the force $\vec{F}$  - acceleration $\vec{a}$ in some inertial frame}
was found with the simplest relation $\vec{a}=\frac{\vec{F}}{m}$ to the mass $m$ of the body. The consistent formulation of mechanical laws has been realized in Galilean  inertial systems. The class of the inertial systems contains (by a convention) the one unique inertial system - the system of remote stars and any reference frame moving with constant velocity relative these remote stars. Then, on the abstract mathematical level arose a ``space" - the linear Euclidean space with appropriate vector operations on forces, momenta, velocities, etc. General relativity and new
astronomical observations concerning accelerated expansion of Universe show that all these constructions are only a good approximation, at best.

The line of Galileo-Newton-Mach and Einstein (with serious reservations about conception of the ``space") argumentations made accent on some absolute global reference frame associated with the system of remote stars. This point of view looks as absolutely necessary for the classical formulation of the inertia principle itself.
Einstein, however, clearly understood the logical inconsistence of the classical formulation of the inertia principle:
``\textbf{The weakness of the principle of inertia lies in this, that it involves an argument in a circle: a mass moves without acceleration if it is sufficiently far from other bodies; we know that it is sufficiently far from other bodies only by the fact that it moves without acceleration}" \cite{Einstein_1922}. This argument may be repeated with striking force being applied to non-localizable quantum objects since for such objects the ``sufficiently far" distance is simply not defined.  I will use the distance between two pure quantum states as a basic concept instead of the distance between two bodies. Such approach leads to the analysis of eigen-dynamics of
quantum states based on the invariant geometric classification of quantum motions in
$CP(N-1)$ \cite{Le3}.

Up to now the inertia principle has been formulated, say, ``externally", i.e. as if one looks on some body moving relative remote stars or ``space". In such approach only ``mechanical" state of relative motion of bodies expressed by their coordinates in space has been taken into account. However, external force not only changes the inertial
character of its motion:  the body accelerates, moreover -- the body deforms.
The second aspect is especially important for quantum ``particles" since the classical acceleration requires the point-like localization in space-time; such localization is, however, very problematic in the relativistic quantum theory \cite{NW}. ``Standard" QFT
loses the fact that space-time geometry and fundamental dynamical variables are state-dependent \cite{Le1,Le2,LeMa1}.
Therefore space-time itself should be built in the frameworks of a new underling ``quantum geometry".

Body's deformation microscopically means that the deformed body
is already a \emph{different body} with different temperature, etc., since the state of body is changed \cite{Le6}. In the case of inertial motion one has the opposite situation -- the internal state of the body does not change, i.e. body is self-identical during inertial space-time motion. In fact this is the basis of all classical physics.
It is tacitly assumed that all classical objects (frequently represented by material points) are self-identical and they cannot disappear during inertial motion because of the energy-momentum conservation law. The inertia law of Galileo-Newton ascertains this self-conservation ``externally". But objectively this means that \emph{physical state of body (temporary in somewhat indefinite sense) does not depend on the choice of the inertial reference frame}. One may accept this statement as an ``internal" formulation of the inertia law that should be of course formulated mathematically. I put here some plausible reasonings leading to such formulation.

Whereas acceleration serves as geometric
counterpart to the classical interaction (curvature of the world line in Newtonian space
and time is non-zero) there is no adequate geometric counterpart for interaction
in quantum theory. The energy of interaction
expressed by a Hamiltonian $H_{int}$ is an analogue of a classical force. Generally, this interaction leads to the absolute change (deformation) of the quantum state \cite{Le6}. Notice, quantum state is in fact the state of motion \cite{Dirac}. Such motion takes the place in a state space modeled
frequently by some Hilbert space $\mathcal{H}$. But there is no geometric
counterpart to $H_{int}$ in such functional space. However, it is well known that
external force perturbs Goldstone's modes supporting a macroscopic body as a macro-system \cite{TFD}. Thereby, quantum states and their deformations may serve as a ``detector" of the external force or self-force action. Thus, instead of an ``outer" absolute reference
frame
like the system distant stars as \cite{Einstein2} the deformation of quantum state may be used as an ``internal detector" for ``accelerated" space-time motion. The deformation of unlocated pure quantum states refers to the internal degrees of freedom is going in an
internal
``unitary field" that geometrically corresponds to the coset transformations. Therefore
\emph{ coset structure of the quantum state space serves as a new geometric counterpart
of quantum self-interaction} \cite{Le1,Le2,Le6,LeMa1}.

\section{Affine non-Abelian gauge potential}
The origin of inertial forces connected with the inertial mass whose nature is unknown. In connection with this fact it is interesting to note that Schr\"odinger's variational principle leads automatically to the orbital moment term playing the role of centrifugal potential
 \begin{equation}\label{1}
    V_{cf}(r)=\frac{\hbar^2l(l+1)}{2mr^2}
 \end{equation}
acting on the wave function. The non-triviality of Schr\"odinger's variation principle will be especially clear if we compare this principle and its development
\cite{Sch_1} with the initial intuitive postulate of Bohr \cite{Bohr} which simply postulated the counterbalance of the attractive Coulomb force and repulsive centrifugal force acting on rotating point-wise electron
\begin{equation}\label{2}
    \frac{e^2}{r^2}=\frac{mv^2}{r}.
\end{equation}
Second step in this direction was done by Dirac whose relativistic wave equation automatically leads not only to the orbital moment but to the spin moment as well.
The third step may be associated with vast area of gauge structure discovered in classical mechanics of deformable bodies and ``geometric phase" leading mainly to Coriolis terms (see nice review \cite{Littlejohn} and corresponding references therein).
There is a question: is it possible to use some variational principle leading automatically to the dynamical nature of the inertial mass? I will apply the variational principle (in the form of affine parallel transport of the vector field of $SU(N)$ generators defining infinitesimal transformation in the state space)
since the origin of the inertial mass will be associated with proper quantum motion in the quantum state space, not in the space-time. This affine parallel transport closely connected with the non-Abelian gauge invariance.

The local Abelian gauge invariance was ordinary connected with the invariance of the Maxwell equations. Yang-Mills fields serve as local non-Abelian gauge fields in the Standard Model (SM). New type of the non-Abelian gauge fields arises under the conservation
of the LDV's of quantum particles. Say, stability of the solution of characteristic
equations (see below) under the Jacobi geodesic variation in $CP(3)$ ensures the self-conservation of the electron due to parallel transport of the internal quantum energy-momentum \cite{Le1,Le2}. Thereby, one has the coupling electrons by the Jacobi gauge vector fields.

It is interesting that already about 20 years ago so-called ``magnetic Jacobi fields" have been used as the variation of trajectories of classical particles on K\"ahler manifolds \cite{Adachi}. Deformed geodesics treated as trajectory of classical particles in the
Jacobi magnetic fields obey generalized Jacobi equations with additional term depends
on strength of an uniform magnetic field and the sectional curvature of a complex
projective space. Quantum dynamics requires essentially different approach similar to the geometric phase ideology.

The geometric phase is an intrinsic property of the family
of eigenstates. There are in fact a set of local dynamical variables (LDV)
that like the geometric phase intrinsically depends on eigenstates.
For us these are interesting vector fields
$\xi^k(\pi^1_{(j)},...,\pi^{N-1}_{(j)}): CP(N-1)\rightarrow \mathcal{C}$
associated with the reaction of quantum state $\pi^i_{(j)}$ on the action of internal ``unitary field" $\exp(i\epsilon \lambda_{\sigma})$ given by $\Phi^i_{\sigma}$.
Notation is defined in the equation (14)later.
In view of future discussion of infinitesimal unitary transformations, it is useful to compare \emph{velocity} of variation of the Berry's phase
\begin{equation}
\dot{\gamma}_n(t) = -\textbf{A}_n(\textbf{R})\dot{\textbf{R}},
\end{equation}\label{7}
where $\textbf{A}_n(\textbf{R})=\Im<n(\textbf{R})| \nabla_{\textbf{R}} n(\textbf{R})>$
with the affine parallel transport of the vector field $\xi^k(\pi^1,...,\pi^{N-1})$
given by the equations
\begin{equation}
\frac{d \xi^i}{d\tau}=-\Gamma^i_{kl}\xi^k \frac{d \pi^l}{d\tau}.
\end{equation}\label{8}
The parallel transport of Berry is similar but it is not identical to the affine parallel transport. The last one is the fundamental because this agrees with Fubini-Study ``quantum metric tensor" $G_{ik^*}$ in the base manifold $CP(N-1)$.
The affine gauge field given by the connection
\begin{eqnarray}
\Gamma^i_{mn} = \frac{1}{2}G^{ip^*} (\frac{\partial
G_{mp^*}}{\partial \pi^n} + \frac{\partial G_{p^*n}}{\partial
\pi^m}) = -  \frac{\delta^i_m \pi^{n^*} + \delta^i_n \pi^{m^*}}{1+
\sum |\pi^s|^2}.
\end{eqnarray}\label{3}
is of course more close to the Wilczek-Zee
non-Abelian gauge fields \cite{WZ} where the Higgs potential has been replaced by the affine gauge potential (11) whose shape in the case $CP(1)$ is depicted in Fig. 1.
\begin{figure}[h]
\begin{center}
    \includegraphics[width=4in]{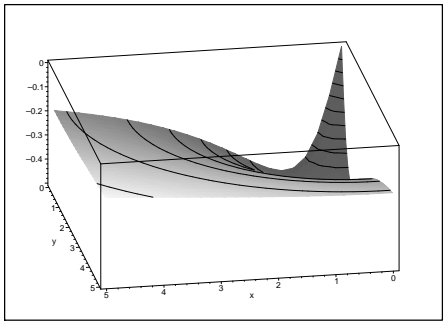}\\
  \caption{The shape of the gauge potential associated with the affine connection in CP(1): $\Gamma=-2\frac{|\pi|}{1+|\pi|^2},
\pi=u+iv. $}\label{fig.1}
  \end{center}
  \end{figure}
It is involved in the affine parallel transport of LDV's \cite{Le4,Le5,Le7} which agrees with the Fubini-Study metric (11).

The transformation law of the connection forms
$\Gamma^i_k = \Gamma^i_{kl} d \pi^l$ in $CP(N-1)$ under the differentiable transformations of local coordinates
$\Lambda^i_m=\frac{\partial \pi^i}{\partial \pi^{'m}}$ is as follows:
\begin{eqnarray}
\Gamma'^i_k = \Lambda^i_m\Gamma^m_j \Lambda^{-1j}_k+d\Lambda^i_s \Lambda^{-1s}_k.
\end{eqnarray}\label{10}
It is similar to the  well known transformations of non-Abelian fields. However the physical sense of these transformations is quite different. Namely: the Cartan's moving reference frame takes here the place of ``flexible quantum setup", whose motion refers to itself with infinitesimally close coordinates. Thus we will be rid of necessity in ``second particle" \cite{AA88} as an external reference frame.

\section{Dynamical space-time }
The distance between two quantum states of electron in $CP(3)$ is
given by the  Fubibi-Study invariant interval $dS_{F.-S.}=G_{ik*}d\pi^i d\pi^{k*}$. The speed of the interval variation is given by the equation
\begin{eqnarray}
(\frac{dS_{F.-S.}}{d\tau})^2=G_{ik*}\frac{d\pi^i}{d\tau}\frac{d\pi^{k*}}{d\tau}
=\frac{c^2}{\hbar^2}G_{ik*}(\Phi^i_{\mu}P^{\mu})(\Phi^{k*}_{\nu}P^{\nu*})
\end{eqnarray}\label{115}
relative ``quantum proper time" $\tau$ where energy-momentum vector field $P^{\mu}(x)$ obeys field equations that will be derived later. This internal dynamics
of \emph{``unlocated quantum states"} in $CP(N-1)$
should be expressed by the \emph{quantum states cum location}
in ``dynamical space-time" coordinates $x^{\mu}$ assuming that variation of
coordinates $\delta x^{\mu}$ arise due to the transformations of Lorentz reference frame that involved in the covariant derivative $\frac{\delta P^{\nu}}{\delta \tau}=  (\frac{\partial P^{\nu}}{\partial x^{\mu}} + \Gamma^{\nu}_{\mu \lambda}P^{\lambda})\frac{\delta x^{\mu}}{\delta \tau}$ in dynamical space-time (DST). Such  procedure may be called ``inverse representation" \cite{Le1,Le2,Le5,LeMa1} since this intended to represent quantum motions in $CP(N-1)$ by ``quantum Lorentz transformation" in DST as it will be described below.

Since there is no a possibility to use classical physical reference frame comprising usual clock and solid scales on the deep quantum level, I will use the ``field frame" from the four components of the vector field of the proper energy-momentum
$P^{\mu}=(\frac{\hbar \omega}{c} ,\hbar \vec{k})$ instead. This means that the period
$T$ and the wave length $\lambda$ of the oscillations associating with an electron's
field are identified with flexible (state-dependent) scales in the DST. Thereby, the local Lorentz ``field frame" is in fact the 4-momentum tetrad
whose components may be locally (in $CP(3)$) adjusted by state dependent
``quantum boosts" and ``quantum rotations".

It is convenient to take Lorentz transformations in the following form
\begin{eqnarray}\label{14}
ct'&=&ct+(\vec{x} \vec{a}_Q) \delta \tau \cr
\vec{x'}&=&\vec{x}+ct\vec{a}_Q \delta \tau
+(\vec{\omega}_Q \times \vec{x}) \delta \tau
\end{eqnarray}
where I put for the parameters of quantum acceleration and rotation the definitions $\vec{a}_Q=(a_1/c,a_2/c,a_3/c), \quad
\vec{\omega}_Q=(\omega_1,\omega_2,\omega_3)$ \cite{G} in order to have for the ``proper quantum time" $\tau$ the physical dimension of time. The expression for the
``4-velocity" $ V^{\mu}$ is as follows
\begin{equation}\label{15}
V^{\mu}_Q=\frac{\delta x^{\mu}}{\delta \tau} = (\vec{x} \vec{a}_Q,
ct\vec{a}_Q  +\vec{\omega}_Q \times \vec{x}).
\end{equation}
The coordinates $x^\mu$ of an imaging point in dynamical space-time serve here merely for the parametrization of the energy-momentum distribution in the ``field
shell'' described by quasi-linear field equations \cite{Le1,LeMa1} that will be derived below.

\section{Self-interacting quantum electron}
Since the spicetime priority is replaced by the priority of the state space,
operators corresponding quantum dynamical variables should be expressed not in the terms space-time coordinates like spatial, temporal differentials or angles but in terms of coordinates of state vectors. Furthermore, deleting redundant common multiplier, one may use local projective coordinates of rays. These local state space coordinates will be arguments of the local dynamical variables represented by the vector fields on $CP(N-1)$ \cite{Le5}.

Further, one needs the invariant classification of quantum motions \cite{Le3}. This invariant classification is the quantum analog of classical conditions of inertial and accelerated motions. They are rooted into the global geometry of the dynamical group manifold. Namely, the geometry of $G=SU(N)$, the isotropy group $H=U(1)\times U(N-1)$ of the pure quantum state and the coset $G/H=SU(N)/S[U(1)\times U(N-1)]$ as geometric
counterpart of the self-interaction, play an essential role in the classification of
quantum motions \cite{Le3}.

In order to formulate the quantum (internal) energy-momentum conservation law in the state space, let us discuss the local eigen-dynamics of quantum system with finite quantum degrees of freedom $N$. It will be realized below in the model of self-interacting quantum electron where spin/charge degrees of freedom in $C^4$ have been taken into account
\cite{Le2}. The LDV's like the energy-momentum and should be expressed in terms of the projective local coordinates $\pi^k, \quad 1 \leq i,k,j \leq N-1$ of quantum state $|\Psi> = \psi^a|a>, \quad 1\leq a \leq N$, where $\psi^a$ is a homogeneous coordinate on CP(N-1)
\begin{equation}
\pi^i_{(j)}=\cases{\frac{\psi^i}{\psi^j},&if $ 1 \leq i < j$ \cr
\frac{\psi^{i+1}}{\psi^j}&if $j \leq i < N$}
\end{equation}\label{1}
since $SU(N)$ acts effectively only on the space of rays,
i.e. on equivalent classes relative the relation of equivalence of quantum states distanced by a non-zero complex multiplier. LDV's will be represented by linear combinations of $SU(N)$ generators in local coordinates of $CP(N-1)$ equipped with the Fubini-Study metric \cite{KN}
\begin{equation}
G_{ik^*} = [(1+ \sum |\pi^s|^2) \delta_{ik}- \pi^{i^*} \pi^k](1+
\sum |\pi^s|^2)^{-2}.
\end{equation}\label{2}
Hence the internal dynamical
variables and their norms should be state-dependent, i.e. local in
the state space \cite{Le4}. These local dynamical variables realize
a non-linear representation of the unitary global $SU(N)$ group in
the Hilbert state space $C^N$. Namely, $N^2-1$ generators of $G =
SU(N)$ may be divided in accordance with the Cartan decomposition:
$[B,B] \in H, [B,H] \in B, [B,B] \in H$. The $(N-1)^2$ generators
\begin{eqnarray}
\Phi_h^i \frac{\partial}{\partial \pi^i}+c.c. \in H,\quad 1 \le h
\le (N-1)^2
\end{eqnarray}\label{3}
of the isotropy group $H = U(1)\times U(N-1)$ of the ray (Cartan
sub-algebra) and $2(N-1)$ generators
\begin{eqnarray}
\Phi_b^i \frac{\partial}{\partial \pi^i} + c.c. \in B, \quad 1 \le b
\le 2(N-1)
\end{eqnarray}\label{4}
are the coset $G/H = SU(N)/S[U(1) \times U(N-1)]$ generators
realizing the breakdown of the $G = SU(N)$ symmetry of the generalized coherent states (GCS's). Here $\Phi^i_{\sigma}, \quad 1 \le \sigma \le N^2-1 $
are the coefficient functions of the generators of the non-linear
$SU(N)$ realization as follows
\begin{equation}
\Phi_{\sigma}^i = \lim_{\epsilon \to 0} \epsilon^{-1}
\biggl\{\frac{[\exp(i\epsilon \hat{\lambda}_{\sigma})]_m^i \psi^m}{[\exp(i
\epsilon \hat{\lambda}_{\sigma})]_m^j \psi^m }-\frac{\psi^i}{\psi^j} \biggr\}=
\lim_{\epsilon \to 0} \epsilon^{-1} \{ \pi^i(\epsilon
\hat{\lambda}_{\sigma}) -\pi^i \}.
\end{equation}\label{5}
Thereby each of the $N^2-1$ generators $\hat{\lambda}_{\sigma}$
may be represented by vector fields $\vec{G}_{\sigma}$ comprising the coefficient
functions $\Phi_{\sigma}^i$ contracted with the corresponding partial derivatives
$\frac{\partial }{\partial \pi^i} = \frac{1}{2}
(\frac{\partial }{\partial \Re{\pi^i}} - i \frac{\partial }{\partial
\Im{\pi^i}})$ and $\frac{\partial }{\partial \pi^{*i}} = \frac{1}{2}
(\frac{\partial }{\partial \Re{\pi^i}} + i \frac{\partial }{\partial
\Im{\pi^i}})$ as follows:
\begin{equation}
\vec{G}_{\sigma} = \Phi_{\sigma}^i \frac{\partial }{\partial \pi^i}+ \Phi_{\sigma}^{*i} \frac{\partial }{\partial \pi^{*i}}.
\end{equation}\label{6}

There are a lot of attempts to build speculative model of electron
as extended compact object in existing space-time, see for example \cite{Dirac2}.
The  model of the extended electron proposed here is quite different.
Self-interacting quantum electron is \emph{a periodic motion of quantum degrees of freedom along closed geodesics} $\gamma$ obeying equation
\begin{equation}
\nabla_{\dot{\gamma}}\dot{\gamma} =0
\end{equation}
in the projective Hilbert state space $CP(3)$. Namely, it is assumed that the motion of spin/charge degrees of freedom comprises of stable attractor in the state space, whereas its ``field-shell" in dynamical space-time arises as a consequence of the local conservation law of the proper energy-momentum vector field. \emph{This conservation law leads to PDE's whose solution give the distribution of energy-momentum in DST that keeps motion of spin/charge degrees of freedom along geodesic in $CP(3)$.}
The periodic motion of quantum spin/charge degrees of freedom generated by
the coset transformations from $G/H=SU(4)/S[U(1)\times U(3)]=CP(3)$
will be associated with inertial ``mechanical mass" and the gauge transformations from $H=U(1)\times U(3)$ rotates closed geodesics in $CP(3)$ as whole. These transformations will be associated with Jacobi fields corresponding mostly to the electromagnetic energy.

In order to built the LDV corresponding to the internal energy-momentum of relativistic quantum electron we shall note that the matrices
\begin{eqnarray}
\hat{\gamma}_0= \left(
\matrix{1&0&0&0 \cr
0&1&0&0 \cr
0&0&-1&0 \cr
0&0&0&-1} \right), \quad
\hat{\gamma}_1= \left(
\matrix{0&0&0&-i \cr
0&0&-i&0 \cr
0&i&0&0 \cr
i&0&0&0} \right),\cr
\hat{\gamma}_2= \left(
\matrix{0&0&0&-1 \cr
0&0&1&0 \cr
0&1&0&0 \cr
-1&0&0&0} \right),
\hat{\gamma}_3= \left(
\matrix{0&0&-i&0 \cr
0&0&0&i \cr
i&0&0&0 \cr
0&-i&0&0} \right),
\end{eqnarray}
originally introduced by Dirac \cite{Dirac_1} may be represented as linear combinations of the ``standard" $SU(4)$ $\lambda$-generators \cite{Close}
\begin{eqnarray}
\hat{\gamma}_0 = \hat{\lambda}_3+\frac{1}{3}[\sqrt{3}\hat{\lambda}_8-
\sqrt{6}\hat{\lambda}_{15}], \quad
\hat{\gamma}_1 = \hat{\lambda}_2+\hat{\lambda}_{14},\cr
\hat{\gamma}_2 = \hat{\lambda}_1-\hat{\lambda}_{13},\quad
\hat{\gamma}_3 = -\hat{\lambda}_5+\hat{\lambda}_{12}.
\end{eqnarray}
Since any state $|S>$ has the isotropy group
$H=U(1)\times U(N)$, only the coset transformations $G/H=SU(N)/S[U(1)
\times U(N-1)]=CP(N-1)$ effectively act in $C^N$. One should remember,
however, that the concrete representation of hermitian matrices
belonging to subsets $h$ or $b$ (as defined above) depends on a priori chosen vector (all ``standard" classification of the traceless matrices of Pauli, Gell-Mann, etc., is based on the vector $(1,0,0,...,0)^T$).   \emph{The Cartan's decomposition of the algebra $AlgSU(N)$ is unitary invariant and I will use it instead of Foldy-Wouthuysen decomposition in ``even" and ``odd" components.}

Infinitesimal variations of the proper energy-momentum evoked by interaction
charge-spin degrees of freedom (implicit in $\hat{\gamma}^{\mu}$ ) that may be
expressed in terms of local coordinates $\pi^i$ since there is a
diffeomorphism between the space of the rays $CP(3)$ and the $SU(4)$
group sub-manifold of the coset transformations
$G/H=SU(4)/S[U(1) \times U(3)]=CP(3)$ and the isotropy group $H=U(1) \times U(3)$
of some state vector. It will be expressed by the combinations
of the $SU(4)$ generators $\hat{\gamma}_{\mu}$
of unitary transformations that will be defined by an equation arising
under infinitesimal variation of the energy-momentum
\begin{equation}
\Phi_{\mu}^i(\pi) = \lim_{\epsilon \to 0} \epsilon^{-1}
\biggl\{\frac{[\exp(i\epsilon \hat{\gamma}_{\mu})]_m^i \psi^m}{[\exp(i
\epsilon \hat{\gamma}_{\mu})]_m^j \psi^m }-\frac{\psi^i}{\psi^j} \biggr\}=
\lim_{\epsilon \to 0} \epsilon^{-1} \{ \pi^i(\epsilon
\hat{\gamma}_{\mu}) -\pi^i \},
\end{equation}\label{14}
arose in a nonlinear local realization of $SU(4)$ \cite{Le2}. Here
$\psi^m, 1\leq m \leq 4$ are the ordinary bi-spinor amplitudes. The twelve coefficient functions  $\Phi_{\mu}^i(\pi)$ in the map $U_1:\{\psi_1 \neq 0\}$ are as follows:
\begin{eqnarray}
\Phi_{0}^1(\pi)&=&0, \quad \Phi_{0}^2(\pi)=-2i\pi^2,
\quad \Phi_{0}^3(\pi)=-2i\pi^3; \cr
\Phi_{1}^1(\pi)&=&\pi^2 -\pi^1 \pi^3,
\quad \Phi_{1}^2(\pi)=-\pi^1 -\pi^2 \pi^3,
\quad \Phi_{1}^3(\pi)=-1 -(\pi^3)^2; \cr
\Phi_{2}^1(\pi)&=&i(\pi^2 +\pi^1 \pi^3),
\quad \Phi_{2}^2(\pi)=i(\pi^1 +\pi^2 \pi^3),
\quad \Phi_{2}^3(\pi)=i(-1 +(\pi^3)^2); \cr
\Phi_{3}^1(\pi)&=&-\pi^3 -\pi^1 \pi^2,
\quad \Phi_{3}^2(\pi)=-1 -(\pi^2)^2,
\Phi_{3}^3(\pi)=\pi^1 -\pi^2 \pi^3.
\end{eqnarray}\label{15}

Now I will define the $\Gamma$-vector field
\begin{equation}
\vec{\Gamma}_{\mu}=\Phi_{\mu}^i(\pi^1,\pi^2,\pi^3)\frac{\partial}{\partial \pi^i}
\end{equation}\label{16}
and then the internal energy-momentum operator will be defined as the \emph{functional
vector field}
\begin{equation}\label{17}
P^{\mu}\vec{\Gamma}_{\mu}\Psi(\pi^1,\pi^2,\pi^3)
= P^{\mu}(x) \Phi_{\mu}^i(\pi^1,\pi^2,\pi^3)
\frac{\partial}{\partial \pi^i}\Psi(\pi^1,\pi^2,\pi^3) + c.c.
\end{equation}
acting on the ``total wave function",
where the ordinary 4-momentum $P^{\mu}=(\frac{E}{c}-\frac{e}{c}\phi,\vec{P} -
\frac{e}{c} \vec{A})=(\frac{\hbar \omega}{c}-\frac{e}{c}\phi,\vec{\hbar k} -
\frac{e}{c} \vec{A})$ (not operator-valued) should be identified with
the solution of quasi-linear ``field-shell" PDE's  for the contravariant
components of the energy-momentum tangent vector field in $CP(3)$
\begin{equation}\label{18}
P^i(x,\pi)=P^{\mu}(x)\Phi_{\mu}^i(\pi^1,\pi^2,\pi^3),
\end{equation}
where $P^{\mu}(x)$ is energy-momentum distribution that comprise of ``field-shell" of the self-interacting electron.

One sees that infinitesimal variation of the internal energy-momentum is represented by
the operator of partial differentiation in complex local coordinates $\pi^i$
with corresponding coefficient functions $\Phi_{\mu}^i(\pi^1,\pi^2,\pi^3)$.
Then the single-component ``total wave function"
$\Psi(\pi^1,\pi^2,\pi^3)$ should be
studied in the framework of new quasi-linear PDE's \cite{Le2,LeMa1}.
There are of course four such
functions $\Psi(\pi^1_{(1)},\pi^2_{(1)},\pi^3_{(1)})$, $\Psi(\pi^1_{(2)},
\pi^2_{(2)},\pi^3_{(2)})$,
$\Psi(\pi^1_{(3)},\pi^2_{(3)},
\pi^3_{(3)}), \Psi(\pi^1_{(4)},\pi^2_{(4)},
\pi^3_{(4)})$
- one function in each local map.

Since the least action principle is correct only in average that is clear
from Feynman's summation of quantum amplitudes a more deep principle should
be used for the derivation of fundamental quantum equations of motion. The quantum formulation of the inertia law has been used \cite{Le1}.
The ``field-shell" equations are derived as the consequence of the
conservation law of the proper energy-momentum \cite{Le2,LeMa1}.

What the inertia law means for quantum system and its states?
Formally the classical inertia principle is tacitly accepted in the package with relativistic invariance. But we already saw that the problem of identification and localization of quantum particles in classical space-time is problematic and therefore they require clarification and mathematically correct formulation. I assumed that quantum version of the inertia law may be formulated as follows:

\textbf{The inertial quantum motion of the quantum system is expressed as a self-conservation
of its local dynamical variables like proper energy-momentum, spin, charge, etc.
in the quantum state space, not in space-time}.

The conservation law of the energy-momentum vector field in $CP(3)$ during
inertial evolution will be expressed by the equation of the affine parallel transport
\begin{equation}
\frac{\delta [P^{\mu}(x) \Phi_{\mu}^i(\pi)]}{\delta \tau}=0,
\end{equation}\label{32}
which is equivalent to the following system of four coupled quasi-linear PDE's
for the dynamical space-time distribution of the energy-momentum ``field-shell" of
the quantum state
\begin{equation}
V^{\mu}_Q (\frac{\partial
P^{\nu}}{\partial x^{\mu} } + \Gamma^{\nu}_{\mu \lambda}P^{\lambda})=
-\frac{c}{\hbar}(\Gamma^m_{mn} \Phi_{\mu}^n(\pi) + \frac{\partial
\Phi_{\mu}^n (\pi)}{\partial \pi^n}) P^{\nu}P^{\mu},
\end{equation}\label{33}
and ordinary differential equations for relative amplitudes giving in fact the definition
of the proper energy-momentum $P^{\mu}$
\begin{equation}
\quad \frac{d\pi^k}{d\tau}= \frac{c}{\hbar}\Phi_{\mu}^k P^{\mu}.
\end{equation}\label{34}
These equations serve as the \emph{equations of characteristics} for the linear ``super-Dirac" equation
\begin{equation}\label{35}
i P^{\mu}\Phi_{\mu}^i(\pi)\frac{\partial \Psi}{\partial \pi^i} + c.c. =mc \Psi
\end{equation}
that agrees with ODE
\begin{equation}\label{36}
i \hbar \frac{d \Psi}{d \tau} =mc^2 \Psi
\end{equation}
for single total state function $\Psi$ of free but self-interacting quantum
electron ``cum location" moving in DST like free material point with the rest mass $m$.

Simple relation given by the Fubini-Study metric for the square of the frequency
associated with the velocity traversing geodesic line
during spin/charge variations in $CP(3)$ sheds the light on the mass problem
of self-interacting electron. Taking into account the ``off-shell" dispersion law
\begin{eqnarray}
\frac{\hbar^2}{c^2}\Omega^2 =\frac{\hbar^2}{c^2} G_{ik*}\frac{d\pi^i}{d\tau}\frac{d\pi^{k*}}{d\tau}
=G_{ik*}(\Phi^i_{\mu}P^{\mu})(\Phi^{k*}_{\nu}P^{\nu*}) \cr
 =(G_{ik*}\Phi^i_{\mu}\Phi^{k*}_{\nu}) P^{\mu} P^{\nu*}
= G_{\mu \nu}P^{\mu}P^{\nu*}=m^2 c^2
\end{eqnarray}\label{115}
one has
\begin{eqnarray}
i \frac{d \Psi}{d \tau} = \frac{mc^2}{\hbar} \Psi = \Psi \sqrt{G_{ik*}\frac{d\pi^i}{d\tau}\frac{d\pi^{k*}}{d\tau}} =\Psi \frac{1}{d\tau} \sqrt{dS^2_{F.-S.}}= \pm \Omega \Psi
\end{eqnarray}\label{116}
and, therefore,
\begin{equation}\label{37}
\Psi(T) = \Psi(0) e^{\pm i S_{F.-S.}}= \Psi(0) e^{\pm i \int_0^T \Omega d\tau} .
\end{equation}
Note that $\Omega(\tau)$ depends on the proper time $\tau$ in our theory.
The metric tensor of the local DST in the vicinity of electron
$G_{\mu \nu} = G_{ik*}\Phi^i_{\mu}\Phi^{k*}_{\nu}$ is state-dependent,  therefore the gravity in the vicinity of the electron generated by the coset
transformations. All this and the general coordinate invariance do not considered
in this paper since the space-time curvature is the second order effect \cite{G}
in comparison with Coriolis contribution to the pseudo-metric
\cite{Littlejohn} in boosted and accelerated
state dependent Lorentz reference frame.

The system of quasi-liner PDE's (25) following from the conservation law
has been shortly discussed under strong simplification assumptions
\cite{Le2,LeMa1}. The theory of such quasi-liner PDE's equations is well known
\cite{CH}. One has the quasi-linear PDE's system with the identical principle part $V^{\mu}_Q$ for which we will build ODE's system of characteristics
\begin{eqnarray}
\frac{d x^{\nu}}{d \tau}&=&V^{\nu}_Q,\cr
\frac{d P^{\nu}}{d \tau}&=&-V^{\mu}_Q
\Gamma^{\nu}_{\mu \lambda}P^{\lambda}-\frac{c}{\hbar}(\Gamma^m_{mn}
\Phi_{\mu}^n(\gamma) + \frac{\partial
\Phi_{\mu}^n (\gamma)}{\partial \pi^n}) P^{\nu}P^{\mu}, \cr
\frac{d\pi^k}{d\tau} &=& \frac{c}{\hbar}\Phi_{\mu}^k P^{\mu}, \cr
\frac{d \Psi}{d \tau} &=& -i\frac{mc^2}{\hbar} \Psi.
\end{eqnarray}\label{44}
In order to provide integration one should to find self-consistent solutions for
``quantum boosts" $\vec{a}$ and ``quantum rotations" $\vec{\omega}$ involved in the ``four velocity" $V^{\mu}_Q$. See for this following paragraph.

\section{Self-consistent problem of quantum boosts and quantum rotations}
Field equations expressing the conservation law of the proper energy-momentum contain state-dependent parameters in quantum boosts and rotations of the Lorentz tetrad. It slides  in DST's depending on a quantum state represented by a point in $CP(3)$. Usually the Lorentz
transformations  $\Omega^{\nu}_{\mu}$ is linear in the free theory and does not depend
on the energy-momentum $P^{\mu}$ but on only the representation of fields. Nonlinearity
of the ``quantum Lorentz transformation" here is a new insight into the self-consistent
problem of state-dependent quantum boosts and quantum rotations.
We will find these parameters from the system of the characteristic equations
\begin{eqnarray}
\frac{dP^{\nu}}{d \tau}&=&-V^{\mu}_Q  \Gamma^{\nu}_{\mu \lambda}P^{\lambda}-\frac{c}{\hbar}(\Gamma^m_{mn} \Phi_{\mu}^n(\pi) + \frac{\partial
\Phi_{\mu}^n (\pi)}{\partial \pi^n}) P^{\nu}P^{\mu}.
\end{eqnarray}
In order to do this one should use equation for the infinitesimal Lorentz transformation of the proper energy-momentum vector  in the following form
\begin{eqnarray}
\frac{dP^{\nu}}{d \tau} =\Omega^{\nu}_{\mu}(P)P^{\mu},
\end{eqnarray}
where
\begin{eqnarray}
\Omega^{\nu}_{\mu}= \left(
\matrix{o&a_1&a_2&a_3 \cr
a_1&0&-\omega_3&\omega_2 \cr
a_2&\omega_3&0&-\omega_1 \cr
a_3&-\omega_2&\omega_1&0} \right),
\end{eqnarray}
with energy-momentum dependent parameters, and, therefore,
\begin{eqnarray}
\Omega^{\nu}_{\mu}(P) P^{\mu} &=&-V^{\mu}_Q  \Gamma^{\nu}_{\mu \lambda}P^{\lambda}-\frac{c}{\hbar}(\Gamma^m_{mn} \Phi_{\mu}^n(\gamma) + \frac{\partial
\Phi_{\mu}^n (\gamma)}{\partial \pi^n}) P^{\nu}P^{\mu}
\end{eqnarray}
or changing the silent index one has
\begin{eqnarray}
\Omega^{\nu}_{\lambda}(P) P^{\lambda} &=&-V^{\mu}_Q  \Gamma^{\nu}_{\mu \lambda}P^{\lambda}-\frac{c}{\hbar}(\Gamma^m_{mn} \Phi_{\lambda}^n(\gamma) + \frac{\partial
\Phi_{\lambda}^n (\gamma)}{\partial \pi^n}) P^{\nu}P^{\lambda}.
\end{eqnarray}
Cancelation of $P^{\lambda}$ leads to the system of algebraic equations for quantum boosts $\vec{a}$ and quantum rotations $\vec{\omega}$
\begin{eqnarray}
\Omega^{\nu}_{\lambda}(P)  &=&-V^{\mu}_Q  \Gamma^{\nu}_{\mu \lambda}-\frac{c}{\hbar}(\Gamma^m_{mn} \Phi_{\lambda}^n(\gamma) + \frac{\partial
\Phi_{\lambda}^n (\gamma)}{\partial \pi^n}) P^{\nu}
\end{eqnarray}
that reads literally as follows:
\begin{eqnarray}
a_1-\frac{a_1(a_1x+a_2y+a_3z)}{c}+\frac{c}{\hbar}L_1P^0 =0 \cr
a_2-\frac{a_2(a_1x+a_2y+a_3z)}{c}+\frac{c}{\hbar}L_2P^0 =0 \cr
a_3-\frac{a_3(a_1x+a_2y+a_3z)}{c}+\frac{c}{\hbar}L_3P^0 =0 \cr
\omega_3+\frac{\omega_3(a_1x+a_2y+a_3z)}{c}-\frac{c}{\hbar} L_2P^1=0 \cr
\omega_1+\frac{\omega_1(a_1x+a_2y+a_3z)}{c}-\frac{c}{\hbar} L_3P^2=0 \cr
\omega_2+\frac{\omega_2(a_1x+a_2y+a_3z)}{c}-\frac{c}{\hbar} L_1P^3=0.
\end{eqnarray}
Their solutions gives quantum proper frequencies and quantum Coriolis-like
accelerations of the co-moving Lorentz reference frame.
\begin{eqnarray}
a_1 &=&c L_1\frac{\hbar \pm \sqrt{\hbar^2+4P^0\hbar(L_1x+L_2y+L_3z)}}{2\hbar(L_1x+L_2y+L_3z)} \cr
a_2 &=&c L_2\frac{\hbar\pm\sqrt{\hbar^2+4P^0\hbar(L_1x+L_2y+L_3z)}}{2\hbar(L_1x+L_2y+L_3z)}               \cr
a_3 &=&c L_3\frac{\hbar\pm\sqrt{\hbar^2+4P^0\hbar(L_1x+L_2y+L_3z)}}{2\hbar(L_1x+L_2y+L_3z)}        \cr
\omega_1 &=& \frac{c L_3 P^2}{\hbar(1+\frac{a_1x+a_2y+a_3z}{c})} \cr
\omega_2 &=& \frac{c L_1 P^3}{\hbar(1+\frac{a_1x+a_2y+a_3z}{c})}\cr
\omega_3 &=& \frac{c L_2 P^1}{\hbar(1+\frac{a_1x+a_2y+a_3z}{c})},
\end{eqnarray}
They have finite limits at the origin of the Lorentz frame $r=0$
\begin{eqnarray}
a_{1lim}=\lim_{r \to 0}{a_1}=\frac{-c L_1 P^0}{\hbar} \cr
a_{2lim}=\lim_{r \to 0}{a_2}=\frac{-c L_2 P^0}{\hbar} \cr
a_{3lim}=\lim_{r \to 0}{a_3}=\frac{-c L_3 P^0}{\hbar} \cr
\omega_{1lim}=\lim_{r \to 0}{\omega_1}= \frac{c L_3 P^2}{\hbar}  \cr
\omega_{2lim}=\lim_{r \to 0}{\omega_2}= \frac{c L_1 P^3}{\hbar} \cr
\omega_{3lim}=\lim_{r \to 0}{\omega_3}= \frac{c L_2 P^1}{\hbar}
\end{eqnarray}
under the choice of the sign ``-" in the expression for $a_{\alpha}$.
Here $L_{\lambda}=\Gamma^m_{mn} \Phi_{\lambda}^n(\pi) + \frac{\partial
\Phi_{\lambda}^n (\pi)}{\partial \pi^n}$  is the divergency of the vector field
of the energy-momentum generator and it is assumed that $\Gamma^{\nu}_{\mu \lambda}$
is the DST connection whose components coincide with boost and rotation instant parameters of the accelerated Lorentz tetrad \cite{G}.

\section{Solutions of the ``field-shell" quasi-linear PDE's}
Let me shortly discuss the system of characteristic equations
\begin{eqnarray}
\frac{dP^{\nu}}{d \tau}&=&-V^{\mu}_Q  \Gamma^{\nu}_{\mu \lambda}P^{\lambda}-\frac{c}{\hbar}(\Gamma^m_{mn} \Phi_{\mu}^n(\gamma) + \frac{\partial
\Phi_{\mu}^n (\gamma)}{\partial \pi^n}) P^{\nu}P^{\mu}.
\end{eqnarray}
of the PDE's system with the identical principle part $V^{\mu}_Q$
\begin{equation}
V^{\mu}_Q \frac{\partial
P^{\nu}}{\partial x^{\mu} } =- V^{\mu}_Q \Gamma^{\nu}_{\mu \lambda}P^{\lambda}
-\frac{c}{\hbar}(\Gamma^m_{mn} \Phi_{\mu}^n(\pi) + \frac{\partial
\Phi_{\mu}^n (\pi)}{\partial \pi^n}) P^{\nu}P^{\mu}.
\end{equation}\label{}
Writing the characteristic equations (32) in symmetrical form with
$S^{\nu}=- V^{\mu}_Q \Gamma^{\nu}_{\mu \lambda}P^{\lambda}
-\frac{c}{\hbar}(\Gamma^m_{mn} \Phi_{\mu}^n(\pi) + \frac{\partial
\Phi_{\mu}^n (\pi)}{\partial \pi^n}) P^{\nu}P^{\mu}$
\begin{eqnarray}
\frac{dx^0}{V^0_Q}=\frac{dx^1}{V^1_Q}=\frac{dx^2}{V^2_Q}=\frac{dx^3}{V^3_Q} \cr
=\frac{dP^0(x)}{S^0}=\frac{dP^1(x)}{S^1}=\frac{dP^2(x)}{S^2}=\frac{dP^3(x)}{S^3} \cr
=\frac{d\pi^1}{\frac{c}{\hbar}P^{\mu}(x)\Phi_{\mu}^1}=\frac{d\pi^2}{\frac{c}
{\hbar}P^{\mu}(x)\Phi_{\mu}^2}=
\frac{d\pi^3}{\frac{c}{\hbar}P^{\mu}(x)\Phi_{\mu}^3}\cr
=\frac{i\hbar d\Psi(x,\pi,P)}{mc^2\Psi(x,\pi,P)}=d\tau
\end{eqnarray}
one may integrate each combination of the equations given above. For example the first pair of equations
\begin{eqnarray}
\frac{dx^0}{V^0_Q}= \frac{dP^0(x)}{S^0}
\end{eqnarray}
may be integrated taking into account $V^0_Q=(\vec{a}\vec{x})$
as follows
\begin{eqnarray}
\int dx^0=ct+cT= \int \frac{V^0_Q}{S^0}dP^0(x).
\end{eqnarray}
The right integral has a form
\begin{eqnarray}
 \int \frac{V^0_Q}{S^0}dP^0(x)=\hbar L_{\alpha}x^{\alpha} \int \frac{\sqrt{1+As}ds}{B (1+\sqrt{1+As})+Cs^2},
\end{eqnarray}
where $s=P^0, \quad A=\frac{4L_{\alpha}x^{\alpha} }{\hbar}, \quad B=-\hbar L_{\alpha}P^{\alpha}, \quad C= 2L_0L_{\alpha}x^{\alpha}$. The substitution $q=\sqrt{1+As}$ leads to the integral from rational integrand as follows
\begin{eqnarray}
 ct+cT= 2A\hbar L_{\alpha}x^{\alpha} \int \frac{q^2 dq}{Cq^4-2Cq^2+A^2Bq+A^2B+C} \cr
 = \hbar L_{\alpha}x^{\alpha}\frac{1}{AB}[2\ln(q+1) \cr
 +\sum_{s=1}^3 \frac{(-R_s^2+R_s(2+(A^2B/C))-1-(A^2B/C)) \ln(q-R_s)}{3R_s^2-2R_s-1}],
\end{eqnarray}
where $R_1, R_2, R_3$ are the roots of the polynomial  $Z^3-Z^2-Z+1+A^2B/C=0$ and
$A^2B/C=-\frac{8L_{\alpha}x^{\alpha}L_{\alpha}P^{\alpha}}{\hbar L_0}$. Therefore
\begin{eqnarray}
\frac{ABc(t+T)}{\hbar L_{\alpha}x^{\alpha} } = \ln(q+1)^2 \cr
+\frac{N_1}{D_1}\ln(q-R_1)+\frac{N_2}{D_2}\ln(q-R_2)+\frac{N_3}{D_3}\ln(q-R_3),
\end{eqnarray}
i.e.
\begin{eqnarray}
\exp(\frac{-4c(t+T)L_{\alpha}P^{\alpha} }{\hbar })\cr
=(q+1)^2 (q-R_1)^{\frac{N_1}{D_1}} (q-R_2)^{\frac{N_2}{D_2}} (q-R_3)^{\frac{N_3}{D_3}},
\end{eqnarray}
where $N_s = -R_s^2+R_s(2+(A^2B/C))-1-(A^2B/C),\quad D_s = 3R_s^2-2R_s-1 $.
One will see that after substitution of these expressions all integrations are elementary
containing rational fractions, logarithms and arctan/arctanh only.
Their shapes are very complicated and the problem of functionally independent invariants did not solved up to now. This solution represents the lump of self-interacting electron in the co-moving Lorentz reference frame.

Up to now we dealt with the field configuration shaping, say, extended self-interaction
electron itself.
``Field-shell" equations gave the distribution of the proper energy-momentum vector field
$P^{\mu}(x)\Phi^i_{\mu}(\pi)$ in the tetrad whose four vectors $P^{\mu}(x)$ are functions over the DST. This means that geodesic motion of spin/charge degrees of freedom have been lifted into the frame fibre bundle over $CP(3)$. No words were told, however, about the interaction between electrons. Next paragraph contains a draft dedicated to possible solution of this problem.

\section{Stability analysis of the characteristic equations and
Jacobi state-dependent gauge fields}
Stationary points of the system of characteristics equations
\begin{eqnarray}
\frac{dP^{\nu}}{d \tau} = \Omega^{\nu}_{\mu}(P) P^{\mu} = 0,
\end{eqnarray}
have been found. Their explicit expression in terms of divergences of
the $SU(4)$ generators fields $L_{\mu}$ presents Coulomb-like 4-potentials
that looks as following:
\begin{eqnarray}
P^0(x)_{st} &=& \frac{1 \pm \sqrt{2}}{2}\frac{\hbar}{L_1x+L_2y+L_3z} \cr
P^1(x)_{st} &=& \pm \frac{\hbar}{2}\frac{L_1((L_2)^2-(L_3)^2)(L_1x+L_2y+L_3z)^{-1}}
{\sqrt{(L_1)^2(L_2)^4+(L_2)^2(L_3)^4+(L_1)^4(L_3)^2 - 3(L_1)^2(L_2)^2(L_3)^2}} \cr P^2(x)_{st} &=& \pm \frac{\hbar}{2} \frac{L_2 ((L_3)^2-(L_1)^2)(L_1x+L_2y+L_3z)^{-1}}
{\sqrt{(L_1)^2(L_2)^4+(L_2)^2(L_3)^4+(L_1)^4(L_3)^2 - 3(L_1)^2(L_2)^2(L_3)^2}}  \cr
P^3(x)_{st} &=& \pm \frac{\hbar}{2} \frac{L_3((L_1)^2-(L_2)^2)(L_1x+L_2y+L_3z)^{-1}}
{\sqrt{(L_1)^2(L_2)^4+(L_2)^2(L_3)^4+(L_1)^4(L_3)^2 - 3(L_1)^2(L_2)^2(L_3)^2}}. \cr
\end{eqnarray}
These components of the stationary energy-momentum vector field being contracted
with $\Phi^i_{\mu}$ serves for direction of some geodesic $\gamma_0$. Then the linear part of deviation from $\gamma_0$ generated by the vector field $p^{\mu}(x)= v^{\mu}(x)e^{\omega \tau}$ involved into the linear analysis of the characteristic stability $\frac{dp^{\mu}(x)}{d\tau}=\hat{M}p^{\mu}(x)$ with the Jacobi matrix
\begin{eqnarray}
\hat{M}=
\left(
\matrix{-\omega+\frac{c}{\hbar}L_{\alpha}P^{\alpha}_{st}&\frac{c}{\hbar}L_1P^0_{st}
&\frac{c}{\hbar}L_2P^0_{st}&\frac{c}{\hbar}L_3P^0_{st} \cr
2\frac{c}{\hbar}L_1P^0_{st}&-\omega+\frac{c}{\hbar}L_2P^2_{st}&\frac{c}{\hbar}L_2P^2_{st}
&-2\frac{c}{\hbar}L_1P^3_{st} \cr
2\frac{c}{\hbar}L_2P^1_{st}&-2\frac{c}{\hbar}L_2P^2_{st}&-\omega+\frac{c}{\hbar}L_3P^3_{st}
&\frac{c}{\hbar}L_3P^2_{st} \cr
2\frac{c}{\hbar}L_3P^0_{st}&\frac{c}{\hbar}L_1P^3_{st}&-2\frac{c}{\hbar}L_3P^2_{st}
&-\omega+\frac{c}{\hbar}L_1P^1_{st}}\right),
\end{eqnarray}
forms the eigenvector problem $M^{\mu}_{\nu}v^{\nu}(x)=\omega v^{\mu}(x)$ for $v^{\mu}(x)$.

The quantum dynamics of spin/charge degrees of freedom of the single self-interacting quantum electron goes along geodesic in $CP(3)$. The lift of this geodesic into the frame fiber bundle leads to the first order PDE's system. It is reasonable assume that interaction of two electrons may deform the geodesic and therefore the lift of the deformed geodesic will be deformed too together with field equations. Adachi and coauthors discussed already so-called Jacobi magnetic fields (closed  K\"ahler 2-form) on K\"ahler manifolds and particularly on $CP(N)$ \cite{Adachi}. One needs, however, the quantum theory of interaction of the extended (non-local) electrons.

I will be concentrated here on the basic variations of the geodesic, namely those that generated by the isotropy group $H=U(1) \times U(3)$. This group is considered as
gauge group transforming one electron motion along geodesic $\gamma_1$ to the motion of second electron along geodesic $\gamma_2$. Thereby these variations may be connected with pure Jacobi vector fields on $CP(3)$.

I will treat the correction for stationary solution (52)
$J^i(x,\pi)=p^{\mu}(x)\Phi_{\mu}^i(\pi)$ as Jacobi vector field, i.e. solution of the
Jacobi equation
\begin{eqnarray}
  \nabla_{\dot{\gamma}}   \nabla_{\dot{\gamma}} J +R(\dot{\gamma},J)\dot{\gamma} = 0.
\end{eqnarray}
This requirement puts additional restriction on the components of anholonomic frame $\Phi_{\mu}^i(\pi)$ in $CP(3)$ that obey to the Duffing type equation with cubic
non-linearity
\begin{eqnarray}
\ddot{\Phi}_{\mu}^i + 2 \omega \dot{\Phi}_{\mu}^i + \omega^2 \Phi_{\mu}^i=
-R^i_{klm*} \Phi_{\mu}^l \dot{\pi}^k \dot{\pi}^{m*}\cr =
-\frac{c^2}{\hbar^2}R^i_{klm*} \Phi_{\mu}^l \Phi_{\nu}^k P^{\nu} \Phi_{\lambda}^{m*} P^{\lambda*},
\end{eqnarray}
where $R^i_{klm*}=\delta^i_k G_{lm*}+\delta^i_l G_{km*}$ is the curvature tensor
of the $CP(3)$.

Complicated equations for $\Phi_{\mu}^i$ requires detailed investigation but solution of the Jacobi equation for $J^i(x,\pi)=p^{\mu}(x)\Phi_{\mu}^i(\pi)$ is well known and this is very easy \cite{Besse}.One may distinguish two kinds of Jacobi fields: tangent Jacobi
vector field $J_{tang}(\pi)=(a_i\tau + b_i)U^i(\pi)$ giving initial frequencies traversing the geodesic and the initial phases, and the normal Jacobi vector field $J_{norm}(\pi)=[c_i \sin(\sqrt{\kappa}\tau)+d_i \cos(\sqrt{\kappa}\tau)]U^i(\pi)$ showing deviation from one geodesic to another \cite{Besse}. There are of course the continuum forms of the geodesic variations but for us only \emph{internal gauge fields} are interesting. We will use only narrow class of such variations: geodesic to geodesic.
It is well known that the isotropy group $H=U(1)\times U(N-1)$ rotates geodesic \cite{KN} whose generators with corresponding
coefficient functions $\Phi_h^i$ may be identified with the normal Jacobi vector fields.
Thereby, two invariantly separated motions of quantum state have been taken into account:

1) \emph{along a tangent Jacobi vector field}, i.e. ``free motion" of spin/charge degrees of freedom along the geodesic line in CP(3) generated by the coset transformations $G/H=SU(4)/S[U(1) \times U(3)] = CP(3)$ (oscillation of a massive mode in the vicinity of a minimum of the affine gauge potential across its valley) and,

2) deviation of geodesic motion \emph{in the direction of the normal Jacobi vector field transversal to the reference geodesic} generated by the isotropy group $H=U(1) \times U(3)$ (oscillation of the massless mode along the valley of the affine gauge potential) .

These oscillators cannot be of course identified with the Fourier components oscillators
of a pure electromagnetic field. But this deformation of geodesic in the base manifold $CP(3)$ induces the nine-parameter deformation of the ``field-shell" quasi-linear PDE's,
described by the Jacobi fields in $CP(3)$ playing the role of non-Abelian
electromagnetic-like field carrier of interaction between electrons. Thus it
should contain the fine structure constant $\alpha=\frac{e^2}{\hbar c}$. Then holomorphic sectional curvature of $CP(3)$
assumed in this paper equal ``1" must be somehow related to the $\alpha$. However, the logical way leading to this connection is unclear yet.

\section{Conclusion and Future Outlook}
This theory has the program character and should be treated as preliminary
framework for the fundamental problems of the grand unification.
It is clearly that gigantic volume of work is left for future.
I would like formulate here very basic and general principles applied to the
single quantum relativistic self-interacting electron in the elementary discussion.
The generalization of this theory on different kind of fermions and their interactions
is only under investigation and will not be mentioned presently.

Analysis of the localization problem in QFT and all consequences of its formal
apparatus like divergences, unnecessary particles, etc., shows that we should
have the realistic physical theory. Such theory requires \emph{intrinsic unification}
of quantum principles based on the fundamental concept of quantum states and the principle of relativity ensures the physical equivalence of any conceivable quantum setup. Realization of such program evokes the necessity of the state-dependent affine gauge field in the state space that acquires reliable physical basis under the quantum formulation of the inertia law (self-conservation of local dynamical variables of quantum particle during inertial motion). Representation of such affine gauge field in dynamical space-time has been applied to the relativistic extended self-interacting Dirac's electron \cite{Le1,Le2,LeMa1}.

The Fubini-Study metric in $CP(N-1)$ is the positive definite metric in the base manifold. Whereas the Lorentz metric is the indefinite pseudo-metric $h_{\mu \nu}$
in the ``vertical" sub-space \cite{Littlejohn} generated by the gauge isotropy
sub-group $H=S[U(1)\times U(N-1)]$ in the frame fibre bundle. The metric tensor
$G_{\mu \nu}$, i.e. the gravity in the vicinity of the electron generated by the coset
transformations, and the general coordinate invariance do not considered in this paper
since the space-time curvature is the second order effect \cite{G}
in comparison with Coriolis contribution to the pseudo-metric in boosting and rotating
state dependent Lorentz reference frame. The application to the general relativity is a
future problem.

The second quantization of the gauge fields has not been discussed in this paper. The Hilbert
space has an indefinite metric in the gauge theories. We need the gauge fixing to
quantize the gauge fields. This may carry out in consistent with the fundamental symmetry
$CP(N -1)$. It will be realized by the choice of the section of the frame fiber bundle
giving by the boundary conditions of the ``field-shell" PDE's for the proper
energy-momentum $P^{\mu}$ (25) and the initial conditions for the components of the anholonomic frame $\Phi^i_{\mu}$ for (55). Details of this gauge fixing will be reported elsewhere.

Dirac's relativistic equations of electron saves mass on-shell condition
under the introduction of two internal quantum degrees of freedom by
matrices belonging to the $AlgSU(4)$. This theory perfectly fits to the electron in the
external electromagnetic field.
The Dirac equation have positive and negative energy solutions, which are
interpreted a particle and an antiparticle. In order to make solutions the positive
energy, we have to consider the Dirac sea or the second quantization. But
in the Section 6, we described the Dirac's single self-interacting quantum
electron where tangent state dependent vector fields to $CP(3)$ replace the Dirac's matrices.
The fields in the vicinity of the electron and their equations of motion have been
derived from the fundamental representation of quantum states motions in the $CP(3)$ geometry. I follow Dirac's ideas \cite{Dirac2} trying to find the mass spectrum of
the electron's generation. This problem is formulated but it does not solved yet.

Usually the Lorentz transformation is linear in the free theory and does
not depend on the energy momentum but on only the representation of fields. Nonlinearity
of the Lorentz transformation may be new insight.

In this paper, only $SU(N)$ group generically connected with
quantum state space geometry is discussed. But, of course, all Lie groups are possible as
gauge groups as it dictated by physical experiments.

\section*{Acknowledgements}
 I am very grateful to Professor T. Adachi for his invitation to ICDG-2012 and interesting discussions. I would like express especial gratitude to the referee for his big efforts for the clarification of some dark places of my presentation and correction of my English.

\end{document}